\documentstyle[11pt,newpasp,twoside,epsf]{article}
\def\edcomment#1{\iffalse\marginpar{\raggedright\sl#1\/}\else\relax\fi}
\reversemarginpar

\begin{document}
\title{Measuring the Size of the Vela Pulsar's Radio Emission Region}
\author{C.R. Gwinn}
\affil{Physics Department, UC Santa Barbara,
Santa Barbara, California, USA}
\author{J.E. Reynolds, D.L. Jauncey}
\affil{Australia Telescope National Facility, Epping, New South Wales, Australia}
\author{H. Hirabayashi, H. Kobayashi, Y. Murata, P.G. Edwards}
\affil{Institute of Space and Astronautical Science, Sagamihara, Kanagawa, Japan}
\author{B. Carlson, S. Dougherty, D. Del Rizzo}
\affil{Dominion Radio Astronomy Observatory, Hertzberg Institute of Astrophysics, 
National Research Council of Canada, Penticton, British Columbia, Canada}
\author{M.C. Britton}
\affil{Swinburne Centre for Astrophysics and Supercomputing, 
Swinburne University of Technology, Hawthorn, Victoria, Australia}
\author{P.M. McCulloch, J.E.J. Lovell}
\affil{Physics Department, University of Tasmania, Hobart, Tasmania, Australia}

\begin{abstract}
We describe the expected distribution 
of intensity for a scintillating source of finite size observed through a scattering
medium, including systematic and instrumental effects.
We describe measurements of the size of the Vela pulsar,
using this technique.
\end{abstract}

\section{Theoretical Background}

Waves from a pointlike source observed through a scattering medium
will suffer random phase changes.  If the phase changes are much
larger than 1 radian, the observer will receive radiation from many
Fresnel zones, and the scattering is said to be ``strong''.  In this
case the electric field at the plane of the observer 
is the sum of the electric field from many lines of sight, 
differing random phases
(Goodman 1985). 
The net electric field is the result of a
random walk.  The electric field is thus drawn from a Gaussian
distribution.  Its square modulus, the intensity, is drawn from an
exponential distribution (Scheuer 1968).

The region from which the observer receives radiation is known as the
scattering disk.  Scattering changes
phases in the Fresnel zones, and thus acts somewhat like
a lens.  If the source is resolved by this ``lens'', the
observed intensity is an incoherent sum from each
part of the source.  
For a source
of small but finite size, the resulting distribution of intensity is the sum of
3 exponentials.  The scales of the smaller exponentials are
approximately the size of the source along either direction on the
sky, in units of the linear resolution of the scattering disk
(Gwinn et al. 1998).  Figure 1 shows
example of the resulting intensity distributions for a point source,
and for a small but resolved source.  When the source is resolved, the
lowest intensities are absent.

\begin{figure}
\plottwo{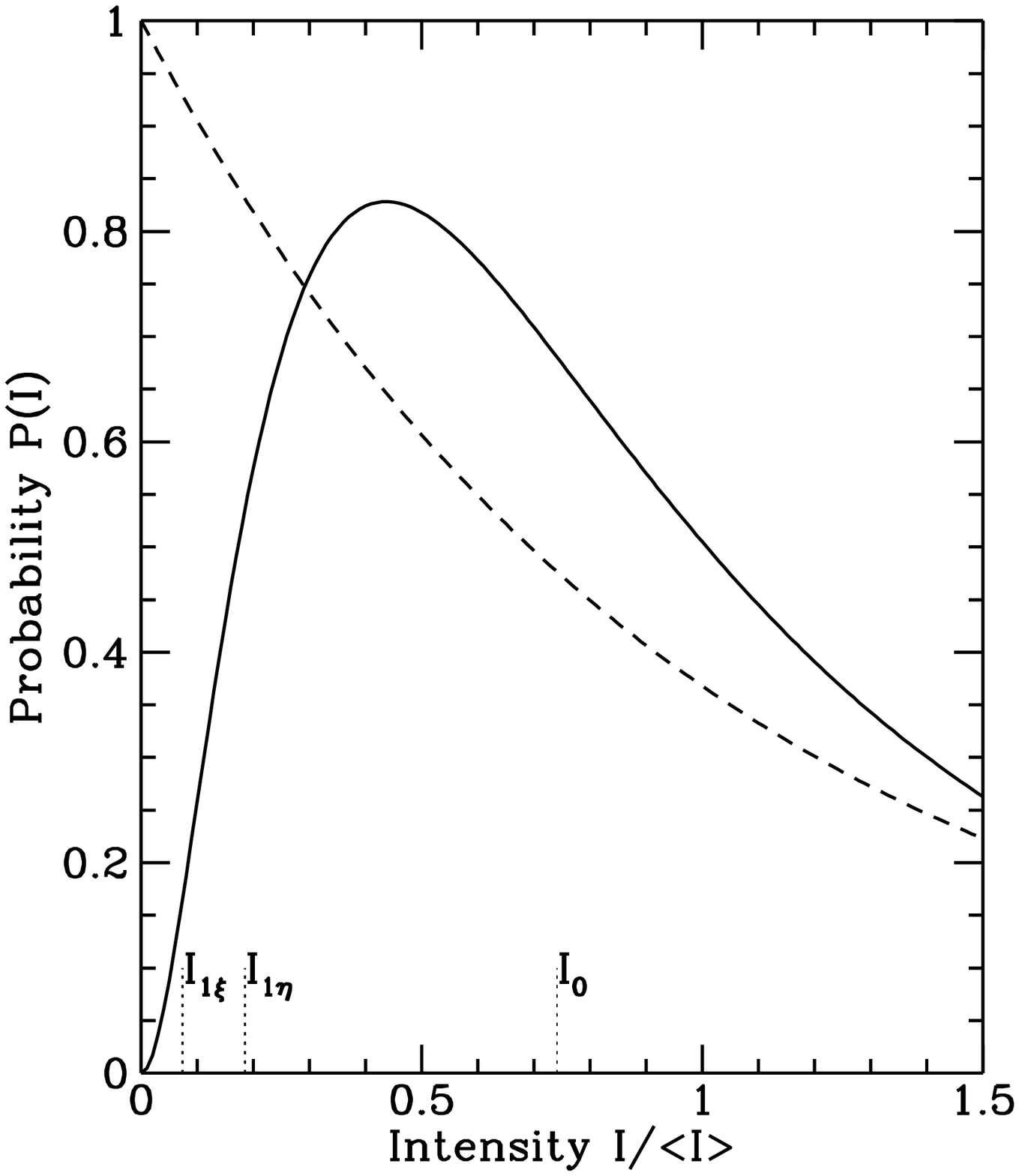}{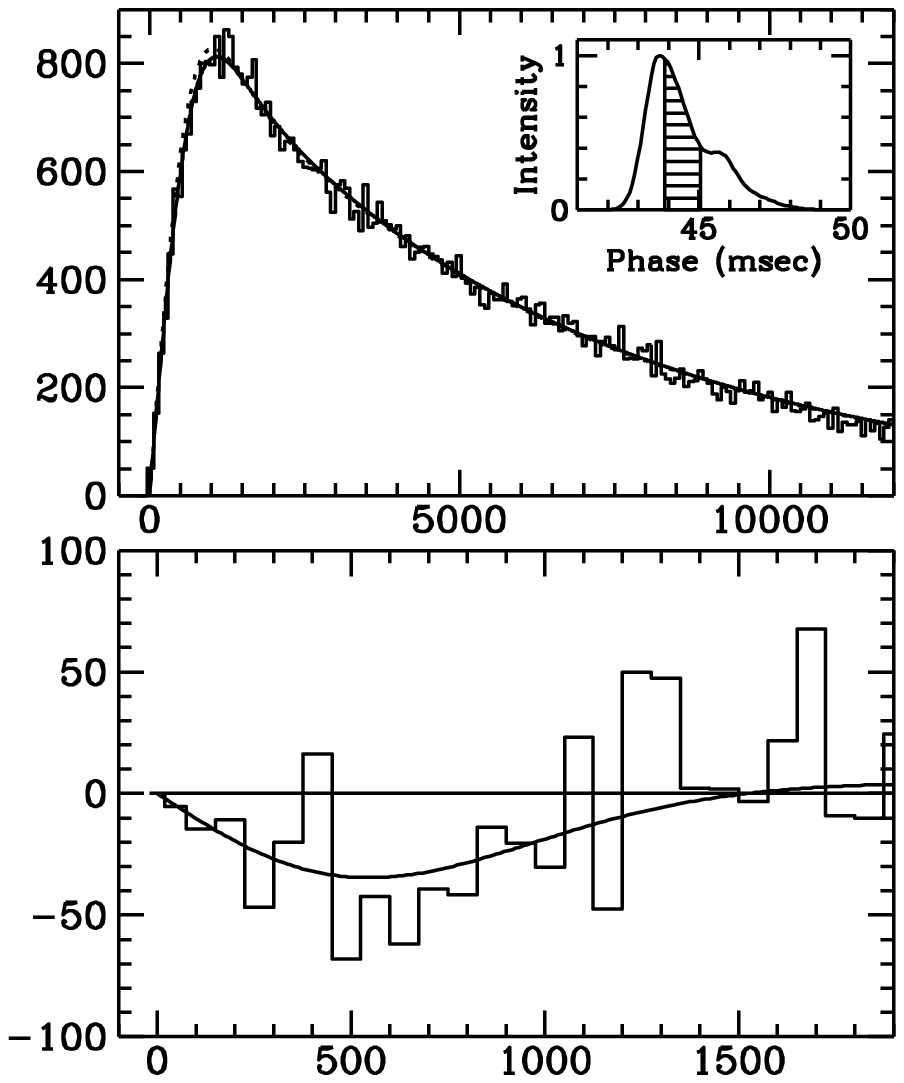}
\caption{Left: Expected distribution of intensity
for a point source in strong scintillation (dashed line);
and for a source of small but finite size (solid line).
From Gwinn et al. (1998).
Right upper: Observed distribution of correlated flux density 
on a short baseline, for 
the Vela pulsar.
Lower: Histogram shows 
residual to the best-fitting distribution for point source,
taking into account the expected noise level. Solid curve
shows best-fitting model including source size.
From Gwinn et al. (2000a).
}
\end{figure}

\section{Observations}

We compare the observed distribution of intensity with theoretical
models to find the size of the Vela pulsar.  The Vela pulsar is a
favorable object for such observations because it is strong and
heavily scattered.  Observations at decimeter wavelengths easily
capture many independent scintles in time and frequency.  We observe
the source interferometrically, rather than with a single dish, to
avoid interference and effects of the substantial
noise baseline seen in single-dish observations.  Details of the
observations are described elsehwere (Gwinn et al. 2000a).

Figure 1 shows an example of the observed distribution of correlated
flux density on the short Tidbinbilla-Parkes baseline for the Vela
pulsar.  
We find a size of $340\pm 80$~km for the data shown in the figure.

Noise affects the distribution shown in Figure 1 strongly.  Like
finite source size, noise reduces the number of points at small
amplitude.  Noise can be measured accurately from observations of
quasars, blank sky, or between pulses.  Its effects can then be
removed.  
The effects of changes in spectral structure on noise from digitization
can also be caculated
(Gwinn et al. 2000b).

Several effects other than noise can also affect the observed
distribution.  Among these are correlator saturation, shot noise,
pulse-to-pulse variability, and gain variations.  These can be either
calculated theoretically, measured from observations, or inferred from
the distribution of intensity. Gwinn et al. (2000a) discuss these
effects in detail.

\section{Modulation Index}

The fact that source size affects the distribution of intensity, in
scintillation, has long been known. (``Stars twinkle, planets do
not.'')  The modulation index, $m=\sqrt{<I^2>-<I>^2}/<I>$, quantifies
the effect (Salpeter 1967, Cohen, Gundermann, \& Harris 1967).
For a point source $m=1$; for an extended source $m<1$,
with smaller modulation $m$ for a larger source,
other factors being equal.
Single-dish observers
used measurements of modulation index to infer source sizes before the
advent of radio interferometry, and 
this technique remains standard at low frequencies
(Hewish, Readhead, \& Duffett-Smith 1974, Hajivasiliou 1992).  
However, it is more subject to scintillation shot
noise, and less immune to systematic effects, than a direct comparison
of distribution functions.

A finite observation necessarily samples a finite number of scintles.
Averages over this sample approximate the statistical
averages $<I^2>$ and $<I>$.
Because the nearly-exponential distribution falls off rapidly
at high intensity, these sums (particularly $<I^2>$) are dominated by 
the relatively rare scintles with the highest intensities.
On the other hand, the effects of source structure are
most important at the lowest intensities,
where the number of scintillations is large,
but the contribution to $<I>$ and $<I^2>$ is small.
Thus, direct estimation of the modulation index
is relatively insensitive to source size and
relatively more sensitive to scintle shot noise
than a direct comparison of the forms of distribution functions.

Correlator saturation also affects the modulation index
strongly, because its effects are largest at high intensity.
Moreover, since the observable is a single number,
rather than a distribution,
it is more difficult to know what effects are playing signficant roles.

Interestingly, Roberts \& Ables (19)
measured the modulation index, as well as the characteristic
time and frequency scales of scintillation,
in their classic study of scattering of southern-hemisphere pulsars.
They report a modulation index of $0.97\pm 0.03$ for the Vela pulsar
at 18~cm wavelength, and of $0.90\pm 0.02$ at
9~cm wavelength.
Interpolation between these values is consistent with our results
quoted above.

Interestingly, Roberts \& Ables find that the modulation index
is smaller at shorter observing wavelengths,
suggesting that the source size is greater.
This conclusion is surprising from the standpoint
of the standard radius-to-frequency mapping.
(Note, however,
that these measurements are
of size rather than emission height.)
The larger inferred size
might reflect on the more complicated pulse
profile of this pulsar at shorter wavelengths (Kern et al. 2000).
On the other hand, it might also reflect systematic effects;
at short wavelengths the scintles have wide bandwidths but
the source remains quite strong, so that correlator saturation should
become more serious.
In contrast, self-noise
and gain variations might be expected to be more important at lower
frequencies. 
Observations of the full distribution of intensity in scintillation,
as a function of wavelength, should indicate the origin of
this variation of modulation index.


\begin{references}
\reference Cohen, M.H., Gundermann, E.J., \& Harris, D.E. 1967, ApJ, 150, 767
\reference Goodman, J.W. 1985, Statistical Optics,
New York: Wiley
\reference Gwinn, C.R., Britton, M.C., Reynolds,
J.E., Jauncey, D.L., King, E.A., McCulloch, P.M., Lovell, J.E.J., \&
Preston, R.A. 1998, ApJ, 505, 928
\reference Gwinn, C.R., Britton, M.C.,
Reynolds, J.E., Jauncey, D.L., King, E.A., McCulloch, P.M., Lovell,
J.E.J., Flanagan, C.S., \& Preston, R.A. 2000, ApJ, in press
\reference Gwinn, C.R., Britton, M.C.,
Carlson, B., Dougherty, S., Del Rizzo, D.,
Reynolds, J.E., Jauncey, D.L., McCulloch, P.M., 
Hirabayashi, H., Kobayashi, H., Murata, Y., \& Edwards, P.G.
2000, in preparation
\reference Hajivassiliou, C.A. 1992, Nature, 355, 232
\reference Hewish, A., Readhead, A.C.S., \& Duffett-Smith, P.J., 1974, Nature, 252, 657
\reference Kern, J., Hankins, T., \& Rankin, J. 2000, these proceedings
\reference Roberts, J.A., \& Ables, J.G. 1982, MNRAS, 201, 1119
\reference Salpeter, E.E. 1967, ApJ, 147, 433
\reference Scheuer, P.A.G. 1968, Nature, 218, 920
\end{references}
\end{document}